\documentclass[twocolumn,superscriptaddress,aps,showpacs,amsmath,footinbib,bibnotes]{revtex4-1}
\pdfoutput=1
\usepackage{graphicx}
\usepackage{amssymb,amsmath}
\usepackage{textcomp} 
\usepackage[bold]{hhtensor}
\usepackage{natbib}
\usepackage{lipsum}
\usepackage{bm}
\usepackage{hyperref}
\usepackage{multirow}
\usepackage{float}
\usepackage{amsmath}
\usepackage{amssymb}
\usepackage{stmaryrd}
\usepackage{mathrsfs}
\usepackage[dvipsnames]{xcolor}
\hypersetup{
	colorlinks,
	linkcolor={red!80!black},
	citecolor={Blue},
	urlcolor={MidnightBlue}
}

\begin{document}

\author{Cheng Wang}
\thanks{These authors contributed equally to this work}
\affiliation{John A. Paulson School of Engineering and Applied Sciences, Harvard University, Cambridge, Massachusetts 02138, USA}
\affiliation{Department of Electronic Engineering, City University of Hong Kong, Kowloon, Hong Kong, China}
\author{Carsten Langrock}
\thanks{These authors contributed equally to this work}
\affiliation{E. L. Ginzton Laboratory, Stanford University, Stanford, CA 94305, USA}
\author{Alireza Marandi}
\affiliation{E. L. Ginzton Laboratory, Stanford University, Stanford, CA 94305, USA}
\affiliation{Department of Electrical Engineering, California Institute of Technology, Pasadena, CA 91125, USA}
\author{Marc Jankowski}
\affiliation{E. L. Ginzton Laboratory, Stanford University, Stanford, CA 94305, USA}
\affiliation{Department of Electrical Engineering and Computer Science, Howard University, Washington DC 20059, USA}
\author{Mian Zhang}
\affiliation{John A. Paulson School of Engineering and Applied Sciences, Harvard University, Cambridge, Massachusetts 02138, USA}
\author{Boris Desiatov}
\affiliation{John A. Paulson School of Engineering and Applied Sciences, Harvard University, Cambridge, Massachusetts 02138, USA}
\author{Martin M. Fejer}\email{fejer@stanford.edu}
\affiliation{E. L. Ginzton Laboratory, Stanford University, Stanford, CA 94305, USA}
\author{Marko Loncar}\email{loncar@seas.harvard.edu}
\affiliation{John A. Paulson School of Engineering and Applied Sciences, Harvard University, Cambridge, Massachusetts 02138, USA}

\date{\today}
\title{Ultrahigh-efficiency second-harmonic generation in nanophotonic PPLN waveguides}

\begin{abstract}
Periodically poled lithium niobate (PPLN) waveguide is a powerful platform for efficient wavelength conversion. Conventional PPLN converters however typically require long device lengths and high pump powers due to the limited nonlinear interaction strength. Here we use a nanostructured PPLN waveguides to demonstrate an ultrahigh normalized efficiency of 2600\%/W-cm$^2$ for second-harmonic generation of 1.5-$\mu$m radiation, more than 20 times higher than that in state-of-the-art diffused waveguides. This is achieved by a combination of sub-wavelength optical confinement and high-fidelity periodic poling at a first-order poling period of 4 $\mu$m. Our highly integrated PPLN waveguides are promising for future chip-scale integration of classical and quantum photonic systems.  
\end{abstract}
\maketitle

The second-order nonlinearity ($\chi^{(2)}$) is responsible for many important processes in modern optics including second-harmonic generation (SHG) and sum-/difference-frequency generation (SFG/DFG) \cite{boyd_nonlinear_2008}. Efficient and compact $\chi^{(2)}$ wavelength converters are crucial elements for a range of applications including entangled photon-pair generation \cite{bouwmeester_experimental_1997}, quantum frequency conversion \cite{zaske_visible--telecom_2012}, low-threshold optical parametric oscillators \cite{leindecker_broadband_2011} and supercontinuum generation \cite{phillips_supercontinuum_2011}. These devices are typically realized using periodically poled lithium niobate (PPLN) crystals, where the periodic domain inversion allows for a quasi-phase-matched (QPM) wavelength conversion process \cite{langrock_all-optical_2006}, and lithium niobate (LiNbO$_3$, LN) provides large $\chi^{(2)}$ nonlinear coefficients (e.g. $d_{33}$ = 25 pm/V) \cite{shoji_absolute_1997}. Since the nonlinear interaction strength is proportional to the light intensity inside the device, using waveguides with tight optical confinement can dramatically increase the conversion efficiencies. Conventional, discrete-component PPLN waveguides however are often based on reverse-proton exchange (RPE) or similar waveguiding technologies which only provide small core-to-cladding index contrast ($\Delta n \sim$ 0.02). As a consequence, the optical mode sizes in these waveguides are large, resulting in low nonlinear interaction strengths, large bending radii in the millimeter range, and overall dispersion properties dominated by that of bulk LN \cite{korkishko_reverse_1998,lim_second-harmonic_1989}. Such devices therefore require long interaction lengths to achieve high conversion efficiencies, making dense chip-scale integration challenging. Furthermore, the lack of waveguide dispersion imposes strict limitation on many applications such as supercontinuum generation \cite{phillips_supercontinuum_2011} and short-pulse optical parametric amplification \cite{leindecker_broadband_2011}.

In recent years, thin-film LN wavelength conversion devices have been pursued to achieve integrated devices with better light confinement and potentially much higher conversion efficiencies \cite{geiss_fabrication_2015,chang_thin_2016,rao_second-harmonic_2016,wang_second_2017,li_broadband_2017,luo_highly-tunable_2018,wang_integrated_2014,lin_second_2015,luo_-chip_2017,wolf_cascaded_2017,wu_lithium_2018,wolf_cascaded_2017,wang_metasurface-assisted_2017}. In this platform, the device layer consists of a single-crystal, sub-micron-thick LN film, which is bonded on top of a low-index substrate (e.g. SiO$_2$) \cite{poberaj_lithium_2012}. Using standard semiconductor microfabrication methods, optical waveguides with sub-wavelength modal confinement and propagation losses as low as 0.03 dB/cm at telecommunication wavelengths have been realized \cite{zhang_monolithic_2017}. The significantly increased optical confinement in principle allows for normalized conversion efficiencies to exceed those of RPE PPLN waveguides ($\sim$ 90\%/W-cm$^2$ for SHG at 1550 nm) by more than an order of magnitude. However, the conversion efficiencies of thin-film LN waveguide devices to date have not been able to provide the promised enhancement, possibly due to a non-ideal overlap between optical modes and/or imperfect periodic poling \cite{geiss_fabrication_2015,chang_thin_2016,rao_second-harmonic_2016,wang_integrated_2014,li_broadband_2017,luo_highly-tunable_2018}. While resonator configurations could be implemented to achieve higher conversion efficiencies, they also lead to compromised operating bandwidths and less tolerance to environmental fluctuations \cite{wang_integrated_2014,lin_second_2015,luo_-chip_2017,wolf_cascaded_2017,wu_lithium_2018,wolf_cascaded_2017}.

In this paper, we demonstrate thin-film PPLN waveguides with ultra-high normalized conversion efficiencies of 2600\%/W-cm$^2$, more than an order of magnitude higher than the previous record in PPLN waveguides while maintaining a large bandwidth \cite{chang_thin_2016}. This is achieved by simultaneously obtaining sub-wavelength optical confinement as well as high-fidelity and repeatable periodic poling. Using a 4-mm-long thin-film LN waveguide, we show continuous-wave SHG from telecom to near-visible wavelengths with a total conversion efficiency of 53\% at an on-chip pump power of 220 mW.
\begin{figure}
	\centering
	\includegraphics[angle=0,width=0.5\textwidth]{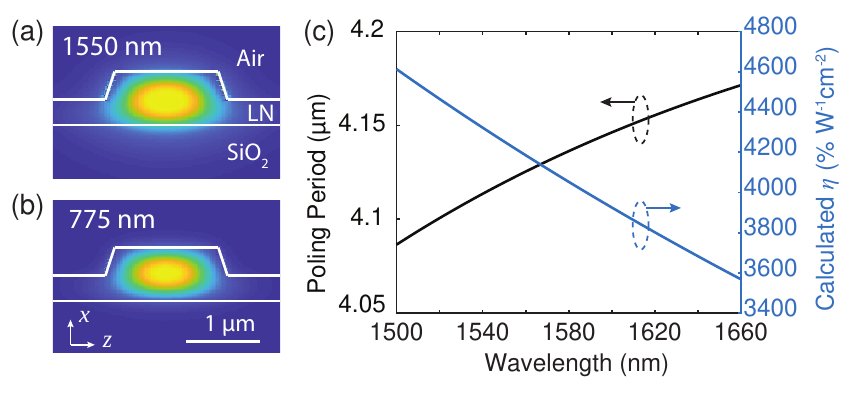}
	\caption{\label{fig1} \textbf{(a,b)} Mode profiles ($E_z$ component) for fundamental TE modes at $\sim$ 1550 nm (a) and $\sim$ 775 nm (b). \textbf{(c)} Numerically calculated poling period for quasi-phase matching (black) and theoretical conversion efficiency (blue) for a typical thin-film PPLN waveguide with a top width of 1440 nm.}
\end{figure}
We achieve the strong nonlinear interaction using first harmonic (FH) and second harmonic (SH) modes that possess high optical confinement and large overlap at the same time. Our x-cut LN ridge waveguides have a device-layer thickness of 600 nm and a waveguide top width of $\sim$ 1400 nm. Figure \ref{fig1}(a-b) shows the numerically simulated optical mode profiles of fundamental transverse-electric (TE$_{00}$) modes at both FH ($\sim$ 1550 nm) and SH ($\sim$ 775 nm) wavelengths, which are used in the conversion process. The TE-polarized modes utilize the highest second-order nonlinear tensor component $d_{33}$ ($d_{zzz}$) in these x-cut films. In order for the wavelength-conversion process to accumulate over a long distance, the well-known phase-matching condition needs to be satisfied, i.e. the momenta of the photons involved in the three-wave mixing process have to be conserved \cite{boyd_nonlinear_2008}. In the case of SHG, this implies that the phase mismatch $\Delta k = 2k_1 – k_2 = 0$, where $k_1$ and $k_2$ are the wavevectors at FH frequency $\omega$ and SH frequency $2\omega$, respectively. In many practical settings, direct phase matching is challenging and limited to certain wavelengths, polarizations and mode combinations. In these circumstances, the QPM method is often used, where the domain orientation of ferroelectric materials like LN is periodically reversed with a period of $\Lambda$ \cite{lim_second-harmonic_1989}. This creates an effective wavevector $k_\textrm{QPM} = 2\pi/\Lambda$, which is used to compensate for the phase mismatch Δk, and to allow monotonic energy transfer over a long interaction length \cite{boyd_nonlinear_2008}. In the case of our thin-film LN waveguides, the required poling period for first-order QPM is $\sim 4~\mu$m, significantly smaller than that in RPE PPLN waveguides due to a much stronger geometric dispersion [Fig. \ref{fig1}(c)]. While an odd multiple of this period could be used for higher-order QPM, it significantly reduces the nonlinear interaction strength \cite{li_broadband_2017}. In this work we focus on first-order periodic poling, which requires precise control over the poling uniformity and yield on a micron scale throughout the chip. 

\begin{figure}
	\centering
	\includegraphics[angle=0,width=0.5\textwidth]{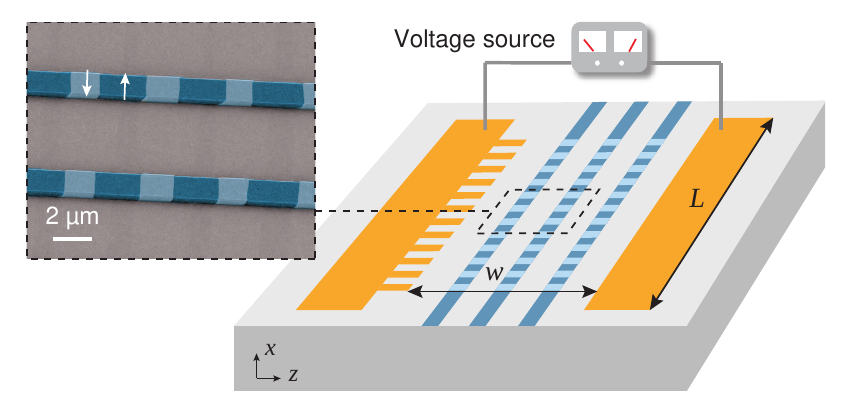}
	\caption{\label{fig2} Schematic of the periodic poling process. Poling finger electrodes are patterned on the surface of an x-cut LN-on-insulator substrate. A high-voltage source is used to periodically reverse the domain orientation of the thin LN film. Ridge waveguides are then fabricated inside the poled region. Inset shows a false-color SEM image of the fabricated waveguide, revealing a poling period of 4.1 $\mu$m with a duty cycle of $\sim$ 39\%.}
\end{figure}

The numerically calculated SHG efficiencies of our nanophotonic PPLN waveguides exceed 4000\%/W-cm$^2$ [Fig. \ref{fig1}(c)], more than 40 times higher than those in RPE waveguides \cite{korkishko_reverse_1998,lim_second-harmonic_1989,sugita_31-efficient_1999,parameswaran_highly_2002,roussev_periodically_2004}. This is due to a strong mode confinement to effective areas of $< 1 \mu$m$^2$ and a large modal overlap between the FH and SH modes [Fig. \ref{fig1}(a-b)]. The normalized conversion efficiency $\eta$ is defined as $P_\textrm{out}/(P_\textrm{in}^2∙L^2)$, where $P_\textrm{in}$ and $P_\textrm{out}$ correspond to FH and SH powers, and $L$ is the device length. We numerically calculate $\eta$ using the following equation:
\begin{equation}
\eta=\frac{2\omega^2d_\textrm{eff}^2}{n_1^2 \epsilon_0 n_2}\cdot\frac{A_{2\omega}}{A_\omega^2}
\label{eq1}
\end{equation}
where $n_1$, $n_2$ are the effective refractive indices of the waveguide modes at FH and SH wavelengths, respectively, $\epsilon_0$ is the vacuum permittivity, $c$ is the speed of light in vacuum. $A_\omega, A_{2\omega}$ are the mode areas at the two wavelengths, defined as:
\begin{equation}
A=\int\textrm{Re}[E_x H_z^*-E_z H_x^*] dxdz
\label{eq2}
\end{equation}
where $E_{x,z}$ and $H_{x,z}$ are the electric and magnetic fields in the corresponding directions, normalized such that the peak electric field is 1. To calculate the effective nonlinear coefficient deff used in Eq. \ref{eq1}, we take into consideration the full nonlinear susceptibility tensor using the following expression:
\begin{equation}
d_\textrm{eff}=\frac{2}{\pi A_{2\omega}}\int\sum_{i,j,k}d_{ijk}E_{i,2\omega}^*E_{j,\omega}E_{k,\omega}dxdz
\label{eq3}
\end{equation}
where $i, j, k \in {x, y, z}.$ According to the symmetry group of LN (class 3m), the majority of these tensor components are zero. The remaining non-vanishing coefficients include: $d_{zzz} = d_{33} = –25$ pm/V, $d_{xzx} = d_{xxz} = d_{yyz} = d_{yzy} = d_{zxx} = d_{zyy} = d_{31} = –4.6$ pm/V and $d_{yyy} = –d_{yxx} = –d_{xxy} = –d_{xyx} = d_{22} = 2.2$ pm/V [1, 7]. From Eq. (1-3) it can be seen that a tight optical confinement (i.e. small mode areas $A_\omega$ and $A_{2\omega}$) dramatically increase the conversion efficiency $\eta$.

\begin{figure}
	\centering
	\includegraphics[angle=0,width=0.5\textwidth]{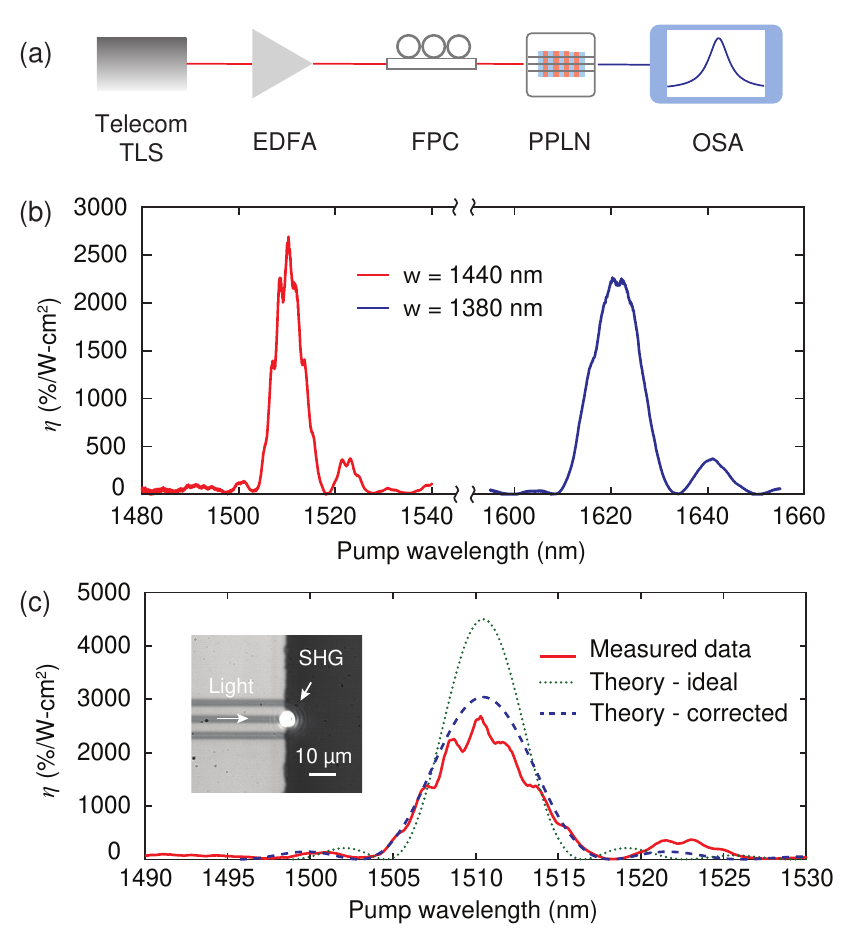}
	\caption{\label{fig3} \textbf{(a)} Schematic of the characterization setup. \textbf{(b)} Measured SHG conversion efficiency versus pump wavelengths for two waveguides with the same poling period but different top widths. A narrower waveguide gives a longer QPM wavelength, consistent with numerical simulation. \textbf{(c)} Zoom-in view of the SHG spectral response of the 1440-nm-wide device (solid curve), together with the theoretically predicted responses (dashed/dotted curves). The green dotted curve assumes a device with ideal poling and structural uniformity, whereas the blue dashed curve shows a corrected transfer function, which takes into account the actual device inhomogeneity. Inset shows a CCD camera image of the scattered SHG light at the output waveguide facet. TLS, tunable laser source; EDFA, erbium doped fiber amplifier; FPC, fiber polarization controller; OSA, optical spectrum analyzer.}
\end{figure}

We demonstrate high-quality first-order periodic poling in thin LN films to enable efficient QPM nonlinear conversion processes (Fig. \ref{fig2}). The devices are fabricated on an x-cut magnesium-oxide- (MgO-) doped LN-on-insulator substrate (NANOLN). We first pattern the poling finger electrodes using a standard photolithography and liftoff process. The metal electrodes consist of a 15-nm Cr adhesion layer and a 150-nm Au layer, deposited by electron-beam evaporation. We perform the periodic domain inversion by applying several 5-ms-long high-voltage pulses at room temperature. The poled region has a width $w = 75 \mu$m and a length $L$ = 4 mm (Fig. \ref{fig2}). Each poled region can accommodate multiple ridge waveguides (3 in our case) without cross-talk due to the strong optical confinement, allowing for dense device integration. After periodic poling, we remove the electrodes using metal etchant. We use aligned electron-beam lithography (EBL) to create waveguide patterns inside the poled regions. The patterns are then transferred to the LN device layer using an optimized Ar$^+$-based dry etching process to form ridge waveguides with smooth sidewalls and low propagation losses \cite{zhang_monolithic_2017}. The inset of Fig. \ref{fig2} shows a false-color scanning electron microscope (SEM) image of the fabricated LN ridge waveguides in the periodically poled region. Here a hydrofluoric acid (HF) wet etching process is used after waveguide formation to examine the periodic poling quality. At a short poling period of 4.1 $\mu$m, the devices still exhibit high poling fidelity with a duty cycle of $\sim 39$\%, essential for achieving high conversion efficiencies. 

We measure normalized SHG efficiencies up to 2600\%/W-cm$^2$ using an end-fire coupling setup shown in Fig. \ref{fig3}(a). Pump light from a continuous-wave telecom tunable laser source (Santec TSL-510, 1480 – 1680 nm) is coupled into the waveguides using a lensed fiber. An in-line fiber polarization controller is used to ensure TE polarization at the input. When the pump laser is tuned to the QPM wavelength, strong scattered SH light can be observed at the waveguide output facet using a CCD camera from top of the chip, as is shown in the inset of Fig. \ref{fig3}(c). After passing through the waveguide, the generated SH light is collected using a second lensed fiber and sent to a visible photodetector (EO Systems) for further analysis. Figure \ref{fig3}(b) shows the measured SHG responses of two waveguides with the same poling period and slightly different top widths of 1440 nm and 1380 nm. Here the fiber-to-chip coupling losses of $\sim$ 10 dB/facet have been calibrated and extracted by measuring the linear optical transmission at both FH and SH wavelengths. The difference between coupling efficiencies of the input and output facets have been de-embedded by comparing the SHG efficiencies when pumping from both sides of the waveguide. A narrower waveguide leads to a longer QPM wavelength due to modal dispersion, as is expected from numerical simulations. The measured η for the 1440- and the 1380-nm-wide devices are 2600\%/W-cm$^2$ and 2300\%/W-cm$^2$, respectively, over an order of magnitude higher than the best values reported in previous PPLN waveguides \cite{lim_second-harmonic_1989,sugita_31-efficient_1999,parameswaran_highly_2002,roussev_periodically_2004,geiss_fabrication_2015,chang_thin_2016,rao_second-harmonic_2016,wang_second_2017,li_broadband_2017,luo_highly-tunable_2018}. 

\begin{figure}
	\centering
	\includegraphics[angle=0,width=0.5\textwidth]{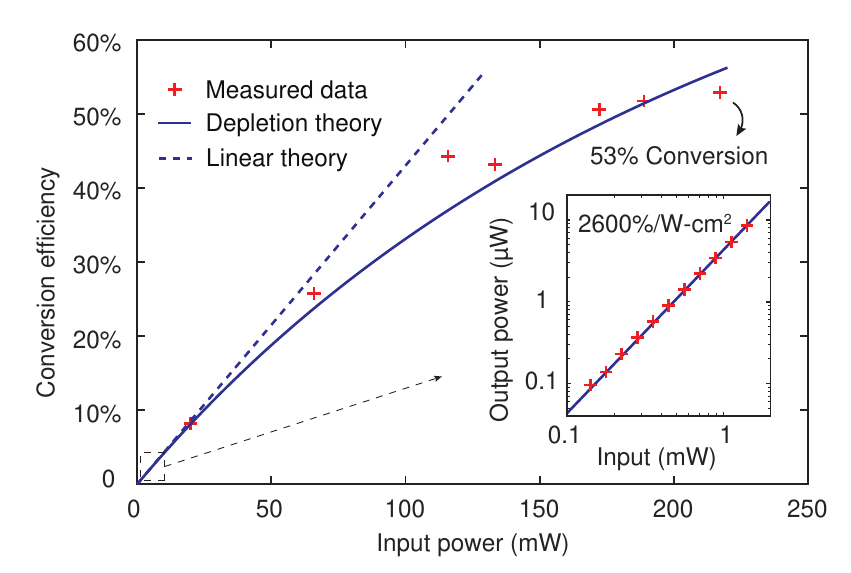}
	\caption{\label{fig4} SHG total conversion efficiency as a function of input power in the pump depletion region, showing a highest measured conversion efficiency of 53\% at an on-chip pump power of 220 mW. Inset shows the input-output power relation in the low-conversion limit.}
\end{figure}

We show that the measured SHG spectral response and conversion efficiency can be well explained using a corrected transfer function model. Figure \ref{fig3}(c) shows a zoom-in view of the SHG response of the 1440-nm-wide device. The green dotted curve corresponds to the theoretically calculated transfer function in the ideal case, showing a sinc-function profile with a maximum normalized conversion efficiency of 4500\%/W-cm$^2$. Compared with the ideal case, the measured SHG response shows a slightly broadened transfer function with a lowered maximum efficiency, likely due to inhomogeneity of the thin-film thickness throughout the 4-mm-long waveguide. While a full characterization of device inhomogeneity would require a phase-sensitive measurement \cite{chang_complex-transfer-function_2014}, we can verify that such inhomogeneity is responsible for the discrepancy between theory and experiment. Since in the absence of pump depletion, inhomogeneous broadening conserves the area of the transfer function, we compare the areas under the measured and calculated transfer functions over the laser tuning bandwidth, which yields a correction factor of 1.28. Using this number, we obtain a corrected transfer function, shown as the blue dashed curve in Fig. \ref{fig3}(c). The corrected transfer function also takes into account the effect from the actual poling duty cycle of $\sim$ 39\%, which reduces the actual $d_\textrm{eff}$ by 7\%. After these corrections, the calculated transfer function shows good agreement with the measured curve in terms of maximum efficiency, QPM bandwidth, as well as side lobe position and shape. The remaining discrepancy between theory and experiment could be attributed to waveguide losses and unoptimized input polarization. 

Furthermore, we show that our nanostructured waveguides could overcome the traditional efficiency-bandwidth trade-off in bulk LN devices thanks to a reduced group-velocity mismatch (GVM). The strong agreement between the measured and theoretically calculated transfer functions [Fig. \ref{fig3}(c)] suggests that the QPM bandwidth in our devices is dominated by waveguide dispersion rather than device inhomogeneity. In this case, the QPM bandwidth is given by $\Delta\lambda=\frac{\lambda^2}{|\Delta\beta'|cL}$, where $\Delta\beta'=(v_{g,2\omega}^{-1}-v_{g,\omega}^{-1})^{-1}$ is the GVM between the FH and SH optical modes. From the experimental results we can extract a GVM of 150 fs/mm, consistent with our numerical simulation prediction. The GVM value in our thin-film waveguides is half of that in bulk LN ($\sim$ 300 fs/mm), resulting in a doubled QPM bandwidth compared with RPE PPLN devices with the same length. Considering the much higher normalized conversion efficiency, our devices show a 50-fold improvement in the efficiency-bandwidth product over RPE waveguides. Further engineering the dispersion property of these thin-film LN waveguides \cite{he_dispersion_2018} could lead to devices with even broader QPM bandwidths.

Finally, we observe the onset of pump depletion and SHG saturation at moderate input powers ($\sim$ 100 mW) in the 4-mm device (Fig. \ref{fig4}), verifying our extracted conversion efficiencies and demonstrating the high-power handling capability of our platform. We use an erbium-doped fiber amplifier (EDFA, Amonics) to further increase the optical power from the pump laser. At the output end, we use an optical spectrum analyzer (OSA, Yokogawa) to simultaneously monitor the optical power at FH and SH wavelengths to measure the pump depletion ratio. The highest measured absolute conversion efficiency in our devices is 53\%, corresponding to the generation of $\sim$ 117 mW at 775 nm in the waveguide using a pump power of 220 mW (Fig. \ref{fig4}). At a high SH optical intensity of $\sim$ 10 MW/cm$^2$ inside the waveguide, we do not observe photorefractive damage of the device after many hours of optical pumping. The measured pump-depletion behavior matches well with the theoretical prediction based on the measured normalized conversion efficiency in the low-conversion limit. The measured input-output power relation in the low-conversion limit follows a quadratic dependence, dictated by the nature of the second-order nonlinear process (inset of Fig. \ref{fig4}).

In conclusion, we show that a nanophotonic PPLN device with sub-wavelength optical confinement and high-quality periodic poling can yield normalized conversion efficiencies that are more than an order of magnitude higher than in traditional devices. Further increasing the device length (e.g. 2 - 3 cm) could allow near-unity overall conversion efficiencies with pump powers in the few-mW range, while maintaining a relatively broad QPM bandwidth. Moreover, our nanophotonic PPLN devices, together with low-loss optical waveguides \cite{zhang_monolithic_2017}, tight bends, and high-speed electro-optic interface \cite{wang_nanophotonic_2018} compatible in the same platform, could provide compact, multi-function and highly efficient solutions at low cost for future classical and quantum photonic systems.

Device fabrication was performed in part at the Harvard University Center for Nanoscale Systems (CNS), a member of the National Nanotechnology Coordinated Infrastructure Network (NNCI), which is supported by the National Science Foundation under NSF ECCS award no. 1541959. We acknowledge funding from National Science Foundation (NSF) (ECCS-1609549, ECCS-1609688, EFMA-1741651); ARL Center for Distributed Quantum Information (W911NF-15-2-0067, W911NF-15-2-0060); Harvard University Office of Technology Development (PSE Accelerator Award); City University of Hong Kong Start-up Fund.

\bibliography{reference}

\end{document}